\documentclass[submission,copyright,creativecommons]{eptcs} 
\providecommand{\event}{David Schmidt} 
\usepackage{amsfonts}

\def\mythanks{\thanks{%
 The research leading to these results has received funding from
 the European Union Seventh Framework Programme (FP7/2007-2013)
 under grant agreement no 318337,
 ENTRA - Whole-Systems Energy Transparency.}}
\def\ruc{\institute{Computer Science, Roskilde University\\
Roskilde, Denmark}}
\def\madsr{\email{madsr@ruc.dk}}
\title{Abstract Interpretation as a Programming Language}
\author{Mads Rosendahl\mythanks\ruc\madsr}
\def\titlerunning{Abstract Interpretation as a Programming Language}
\def\authorrunning{Mads Rosendahl}
\def \Int{\mathbb{N}}
\def \Unit{\mathbb{U}}

\def \lemma#1{\par\vspace{-7pt plus 0pt}\par\paragraph{#1}}
\def \moveup{\vspace{-3pt}}
\def \ttt{\small\tt}
\def \sss{\small\def\baselinestretch{0.9}}
\def \strdom{{\ttt strdom}}
\def \intdom{{\ttt intdom}}

\def \opname#1{{\mathop{#1\vrule depth 0pt width 0pt}\nolimits}}
\def \kw#1{\mathrel{\mbox{\bf #1}}}

\def \sm#1{\opname{\mbox{\sf #1}}}
\def \bfsf#1{\textsf{\textbf #1}}
\def \lfp{\sm{lfp}}

\def \bc#1{\opname{\langle#1\rangle}}
\def \NN{{\opname{\bfsf{N}}}}

\let \ggl\[\let\ggr\]
\def \[{\ggl\begin{array}{@{}lllll}}
\def \]{\end{array}\ggr}

\let \power\wp
\let \fun \rightarrow
\let \imp       \Rightarrow

\let \lub       \sqcup   
\let \glb       \sqcap   
\let \Lub       \bigsqcup   

\let \qleq      \sqsubseteq
\let \qgeq      \sqsupseteq
\def \f#1{\mbox{\it#1}\;}

\newcommand{\XVmm  }{\mbox{\hspace*{15mm}}}

\begin{document}
\maketitle

\begin{abstract}

In David Schmidt's PhD work he explored the use of denotational semantics as a programming language. It was part of an effort to not only treat formal semantics as specifications but also as interpreters and input to compiler generators.  The semantics itself can be seen as a program and one may examine different programming styles and ways to represent states. 

Abstract interpretation is primarily a technique for derivation and specification of program analysis. As with denotational semantics we may also view abstract interpretations as programs and examine the implementation. The main focus in this paper is to show that results from higher-order strictness analysis may be used more generally as fixpoint operators for higher-order functions over lattices and thus provide a technique for immediate implementation of a large class of abstract interpretations. Furthermore, it may be seen as a programming paradigm and be used to write programs in a circular style.

\end{abstract}

\section{Introduction}

The transition from a denotational semantics to an interpreter in a functional language is often fairly straightforward and may even be used to check a specification for mistakes.  
An abstract interpretation is often less obvious to realise and the path from a specification to a prototype implementation may require specialised techniques for finding fixpoints in domains. 
One of the obstacles is that semantics are often written in a higher-order style which is not easy to implement using traditional fixpoint iteration techniques. 
There seem to be two major approaches to making generalised systems for abstract interpretations. One can make a system for a specific language which can then be analysed using a selection of abstract interpretations \cite{Bagnara07}. 
In this way one may represent control flow independently of the analysis and typically avoid some higher-order dependencies in the interpretation. 
An alternative approach is to use attribute grammars as a specification language for abstract interpretations \cite{FokkerS09}. 
This provides a more restricted framework for specifying semantics so that dependencies in the semantics are represented as synthesised and inherited attributes and where the semantics can be interpreted using first or second-order fixpoint iteration.

We present a method for demand-driven fixpoint iteration on domains with higher-order functions. The fixpoint operation can be used in the implementation of a large class of abstract interpretations and is not restricted to analyses that can be implemented using first or second-order fixpoint iteration. The technique uses partial function graphs to represent higher-order objects. The main problem in finding fixpoints for higher-order functions is to establish a notion of {\em neededness} so as to restrict the iteration to those parts of the function that may influence the result. The central principle is to instrument a higher-order function with neededness information so that functional arguments can be tabulated for the needed part at the time of the call and, thus, before they are actually needed in the functional. This is done here through a uniform extension of the domain of values with need information. The result is an iteration strategy which will terminate if the base domains are finite. 

The technique is a generalisation of the minimal function graph approach for fixpoint iteration of second-order functionals. For second-order functionals a demand-driven evaluation of the value of the fixpoint for a given argument will tabulate those parts of the fixpoint that may be called, directly or indirectly, from the initial argument. 

For a number of years the implementation of higher-order strictness analysis on non-flat domains \cite{bha:hos,B66,B104} has been a benchmark for how well different techniques perform. In a previous work \cite{jhr:esah} we constructed a strictness analysis for Haskell. It had some very attractive properties both regarding speed and precision. In this paper we extend that work and show that the underlying fixpoint evaluation method is not restricted to strictness analysis but can be used as a more general fixpoint operation on higher-order domains.

The use of a general fixpoint operator makes it possible to explore circularity as a programming paradigm. Functional definitions can be circular in a number of different ways. Functions may be defined recursively in terms of themselves. We may have programs that use circular data structures, and functions may have circular argument dependencies. Circular data structures or circular dependencies can be used as a powerful algorithmic principle in a lazy language. The phrase ``circular programs'' was coined by R. Bird \cite{bird:circ} who showed how several passes over a data structure can be expressed in a single function, provided one may pass parts of the result of the function as an argument to the function.

In the field of attribute grammars, circularity testing plays an important role. The traditional approach is to disallow circular attribute grammars but under certain restrictions it is well-defined and useful to allow circularity. A number of evaluators for circular attribute grammars have been reported in the literature \cite{f:agfix,libag-g579}.

The fixpoint operation is currently being used to make various resource related analyses for embedded software systems, where we both want to change between different languages and analyse different properties.


\section{Higher-order fixpoint iteration}
\label{sec:hofi}

There are a number of challenges and pitfalls in implementing fixpoint iteration for higher-order functions. We will examine a few of these as a motivation for introducing the demand-driven technique presented later in this paper.

With the extension of strictness analysis to higher-order functions and analysis of lazy lists came examples of simple programs that were surprisingly complicated to analyse. An example examined in numerous papers is the $\f{cat}$ function.
\[
\f{foldr}\, []\; b=b
\\
\f{foldr} f\, (a:\f{as})\,b= f\, a (\f{foldr} f\, \f{as}\, b)
\\
\f{cat} l = \f{foldr} \f{append} l\; [ ]
\]
When we analyse the strictness of this function we examine a function in the domain $[[{\bf 4} \to {\bf 4} \to {\bf 4}] \to {\bf 6} \to{\bf 4} \to {\bf 4}]$, where ${\bf 4}$ is a four-element domain and ${\bf 6}$ is a six-element domain, both with a total ordering. The function domain has in the order of $10^6$ elements \cite{hh:front} and a full fixpoint iteration is quite costly if the function has to be tabulated and re-evaluated until stability.

\paragraph{Partial fixpoint iteration.} 

Instead of tabulating a higher-order function for all values one may want to limit the evaluation to fewer arguments and thus work with partial functions. If one is not careful this may lead to a wrong result.

Let $G: (D_1\fun D_2)\fun (D_1\fun D_2)$ be continuous and let $S_i$ be a sequence of subsets of $D_1$. We will now define a sequence of approximations $\phi_i$ as follows
\[  
   \phi_0 &= \lambda x . \bot
\\  \phi_{i+1} &= \lambda x . \kw{if} x\in S_i 
             \kw{then} G\; \phi_{i}\; x\, \kw{else} \phi_{i}\; x
\]
The aim is to find a limit $\phi_n$ which for some subset of $D_1$ should compute the same function as $\lfp\, G$.

Unfortunately the functions $\phi_i$ are not guaranteed to be monotonic, nor will the sequence in general be increasing.

As an example consider the domain $\{0,1,2\}$ with ordering $0\leq 1\leq 2$, and the function:
\[
   G\; \phi\; x &= \phi(\phi(x))\oplus 1
\\  x \oplus 1 &= (x+1)\glb 2
\]
and sets $S_i=\{1\}$. The functions $\phi_i$ are then:
\[\phi_0 &= [0\mapsto 0,1\mapsto 0,2\mapsto 0]
\\\phi_1 &= [0\mapsto 0,1\mapsto 1,2\mapsto 0]
\\\phi_2 &= [0\mapsto 0,1\mapsto 2,2\mapsto 0]
\\\phi_3 &= [0\mapsto 0,1\mapsto 1,2\mapsto 0]
\\\ldots
\]
%
The problem is that function application in general is not monotonic. As a consequence a selective re-evaluation of the function on needed arguments does not necessarily form an increasing chain and could result in non-termination of a fixpoint algorithm.
Further conditions are needed to guarantee such a sequence to approximate the fixpoint. We will explore such methods in section \ref{sec:domfix}.

\paragraph{Closure-based fixpoint iteration.}

An alternative to using partial functions could be to describe functional arguments as closures. In some situations this may work and give simple solutions to examples like the $\f{cat}$ example above
\cite{mr:hcfi,jr:hmfg}. The technique, however, is not generally applicable as a fixpoint iteration technique as it may lead to infinite sequences of closures.

Consider the functions $ g: {\bf 2} \fun ({\bf 2}\fun {\bf 2}) \fun {\bf 2}$
and $m:{\bf 2} \fun ({\bf 2}\fun {\bf 2}) \fun {\bf 2}\fun {\bf 2}$, where ${\bf 2}$ is the two element domain $\{\bot,\top\}$ with $\bot\qleq\top$, and $\sqcup$ and $\sqcap$ as least upper bound and greatest lower bound operations:
\[
g'\; n\; k     &= n \sqcap ((k\; \top) \sqcup (g'\; n\; (m'\; n\; k))) 
\\
m'\; n\; k\; x &= k(n \sqcap x) 
\]
These functions are generated as strictness function for the CPS converted
factorial function.
\[
g\;n\;k   &=\, {\bf if}\; n= 0 \;{\bf then}\; k\; 1 
          \;{\bf else}\; g\, (n-1)\, (m\; n\; k)
\\
m\; n\; k\; x\; &= k(n*x)
\\
\mathit{id}\; x &= x
\\
\mathit{fac}\; x &= g\; x\;\mathit{id}
\]
One approach to fixpoint iteration could be to represent functional values as closures of partial function applications. We will here write such partial applications in angled brackets. 
The call $g'\;\top\;\bc{\f{bot}}$  with $\f{bot}=\lambda x. \bot$ will lead to the following sequence of calls to $g'$:
\[
g'\; \top\; \bc{bot}
\\
g'\; \top\; \bc{m'\;\top\;\bc{bot}}
\\
g'\; \top\; \bc{m'\;\top\;\bc{m'\;\top\;\bc{bot}}}
\\
...
\]
A closure-based semantics may, however, be used as a standard semantics in an abstract interpretation \cite{jr:hmfg} to prove the correctness of a closure analysis. In a similar style minimal function graphs were introduced \cite{jm:mfg} as an instrumented standard semantics to prove the correctness of constant propagation of an applicative language. It was not introduced as a fixpoint iteration technique.


\section{Domains and fixpoints}
\label{sec:domfix}

The theory behind the circular definitions is based on the fixpoint theorems of monotonic maps on complete lattices or continuous maps on chain-complete partially ordered sets. We are interested in computable fixpoints and will therefore impose some extra requirements regarding finite, unique representation and existence of certain extra operations. In general our approach is to separate the sets of values from the domain structure.

\lemma{Domain.}
A domain $D$ is a set with the following properties and operations.
\begin{itemize}

\item $D$ has a partial order $\qleq$. A partial order is a reflexive, antisymmetric and transitive relation.

\item $D$ has a least element $\bot$. Hence $\forall x \in D . \bot \qleq x$.

\item $D$ is chain-complete. For any subset $(x_i)$ where $x_i\in D$ and where $x_i \qleq x_{i+1}$ (i.e. a chain) there exists a least upper bound $\Lub_i x_i$: 
$\forall d. (\forall i . x_i \qleq  d) \imp \Lub_i x_i \qleq d$.

\item Any two elements of $D$ with an upper bound will also have a least upper bound. The least upper bound of two elements $x$ and $y$ in $D$ is written $x\lub y$. 
\end{itemize}
Notice that we do not require general pair-wise least upper bounds to exist. We can therefore have flat domains and other domains without a top element. The last condition is an alternative to a stronger requirement of $D$ being a complete lattice.
We will frequently use domains where not all pairs of elements have a least upper bound. This is the case for flat domains larger than the two-point domain.

Any element of $D$ which is the least upper bound of a chain without being an element of the chain is called a {\em limit point}. 

We will require that all values which are not limit points can be represented in our programming language in a finite and unique way and that there is a total ordering of values for implementation purposes. This ordering does not need to be related in any way to the partial order on the domain. We will return to the implementation issue in section \ref{sec:imp}.

\lemma{Function domain.} Given a set $S$ and a domain $D$. The set $S\fun D$ of all functions from $S$ to $D$ is a domain. The domain has the following structure:
\\
The bottom element is $\lambda x . \bot_D$, where $\bot_D$ is the bottom element in $D$.
\\
The ordering is defined as $f\qleq g \iff \forall x . f(x)\qleq g(x)$.
\\
Assume $f\qleq h$ and $g\qleq h$ for functions $f$, $g$, and $h$, then
$f\lub g = \lambda x.f(x)\lub g(x)$.
\\
Given a chain $(f_i)$, then $\Lub_i f_i = \lambda x . \Lub_i f_i(x)$.

\smallskip

In the remainder of this paper $S\fun D$ will denote this function domain without any implicit requirement of the functions being monotonic or continuous.

\lemma{Monotonicity.}

A function $f$ from a domain $D_1$ to $D_2$ is monotonic if it preserves the order:
$  \forall x,y \in D_1 . x \qleq y \imp f(x) \qleq f(y) $\\
The function $f$ is {\em continuous} if it is monotonic and preserves the least upper bound of chains.
$  \forall (x_i) \in D . f(\Lub_i x_i) = \Lub_i(f(x_i))$

\lemma{Fixpoint iteration.}

The central result in the domain theory is that continuous functions on a domain have a least fixpoint. A similar result exists for monotonic functions on a complete lattice \cite{B306}. The fixpoint can be found as the limit of a chain starting with the bottom element with repeated applications of the function.
\[
   \bot \qleq f(\bot)\qleq f(f(\bot))\qleq f(f(f(\bot))) \qleq ...
   f(\Lub_i f^i(\bot)) = \Lub_i (f (f^i(\bot)))
\]
We will refer to the fixpoint of the function as $\lfp f$ - the least fixpoint of $f$.

\paragraph*{Pseudo-monotonicity.}

If $f$ is a function on non-functional domains then the fixpoint iteration is immediately applicable \cite{apt:cha}. In the second-order case the main complication is that function application in general is not monotonic as illustrated in the first example of section \ref{sec:hofi}. As a consequence a selective re-evaluation of the function on needed arguments does not necessarily form an increasing chain and could result in non-termination of the fixpoint algorithm. Fortunately function application has another property: pseudo-monotonicity \cite{dix:fix} (in \cite{fs:evenfast} called weak monotonicity).

A function $G: (D_1\fun D_2)\fun (D_1\fun D_2)$ is pseudo-monotonic iff
$G$ preserves monotonicity, is continuous for monotonic arguments and
\[
  \forall f,f': D_1 \fun D_2 ,  f \qleq f' .
\\\XVmm f \mbox{ monotonic}  \imp G(f) \qleq G(f')    \land 
\\\XVmm f' \mbox{ monotonic}  \imp G(f) \qleq G(f')   
\]
Function application satisfies this extra condition.

\paragraph*{Function needs.}

Given $G: (D_1\fun D_2)\fun (D_1\fun D_2)$, a function $f: D_1\fun D_2$ and an element $x\in D_1$. The {\em needs} of the evaluation $G f x$ is the set of values in $D_1$ for which the function $f$ will be called during the evaluation. We will write $\NN\, G\, f\, x$ for this set.

Needs is in this way an internal property but may be externalised \cite{mr:mfg} as follows:
given functions $G$, $f$ and a value $x$, 
let $S=\NN\,G\,f\,x$, then
\[
   \forall g. (\forall y\in S. f\,y=g\,y) \imp G\,f\,x =  G\,g\,x
\]
In other words: if a value is not needed then the function will return the same value independently of how the argument was defined for that value.

This externalised view gives a lower bound of needed arguments.

\[
  y \not\in \NN\, G\, f\, x \imp 
    \forall z\in D_2 . G f x = G (f[y\mapsto z]) x
\] 
where 
\[
  f[y\mapsto z]= \lambda x.\kw{if} x=y \kw{then} z \kw{else} f x 
\]
It is safe to define needs as a super set of arguments that are actually required to compute the value.

\lemma{Partial fixpoint iteration.} 

Let $G: (D_1\fun D_2)\fun (D_1\fun D_2)$ be pseudo-monotonic and let $S_i$ be a sequence of subsets of $D_1$. We will now define a sequence of approximations $\phi_i$ as follows
\[  
   \phi_0 &= \lambda x . \bot
\\  \phi_{i+1} &= \lambda x.\kw{if} x\in S_i \kw{then} \phi_{i}\; x \lub G\; \phi_{i}\; x\, \kw{else} \phi_{i}\; x
\]
\\
These functions will form an increasing chain (see below).

\lemma{Chaotic iteration and minimal function graphs.}
This sequence is similar to chaotic fixpoint iteration \cite{cc:77c} where one will require that needed arguments from one iteration appears in the set of values to be evaluated in the next.
A simple example of this is the sequence
\[
  S_{i+1} = \Lub_{x\in S_i}\NN\, G\, \phi_i\, x
\]
This is the approach often referred to as the minimal function graph sequence \cite{jm:mfg}.

\lemma{Stability.} 

Rather than to consider when to re-evaluate the approximations it is easier to examine the situation for a stabilised iteration sequence where for some $n$
\[  
  \forall i> n . S_i = S_n ,\phi_i=\phi_n
\\
  \forall x \in S_n . \NN\, G\, \phi_n\, x \subseteq S_n
\]
If this is the case then we have
\[
   \forall x \in S_n . \phi_n x = (\lfp\, G) x
\]
\bigskip
In the remainder of this section we will assume we have a stabilised iteration sequence for a function $G$ and sets $(S_i)$, and that it stabilised after $n$ iterations.

\lemma{Lemma.}
The sequence $(\phi_i)$ is well-defined and forms an increasing chain.
\lemma{Proof.}
The least upper bound operation is well-defined since $\phi_i\,x\,$ and $G\,\phi_i\,x\;$ are both less than or equal to  $(\lfp\;G) x$.
\\
Define the chain $(\theta_i)$, where
\[
\theta_0 = \lambda x . \bot
\mbox{~~ and ~~}
\theta_{i+1} = G\;\theta_i
\]
Since $G$ preserves monotonicity we know that $\theta_i$ is monotonic and that $\theta_i\qleq\theta_{i+1}$. By induction we prove that $\phi_i\qleq\theta_i$.
\\
The base case $\phi_0\qleq\theta_0$ holds trivially.
\\
Assume for some $i$ that $\phi_i\qleq\theta_i$, and given $x\in S_i$ then
\[
\phi_{i+1}\, x = \phi_i\;x\;\lub\; G\;\phi_i\;x
\]
and
\moveup
\[
  \phi_i\,x \qleq\theta_i\,x \qleq\theta_{i+1}\,x 
\mbox{~~ and ~~}
  G\,\phi_i\,x \qleq G\,\theta_i\,x =\theta_{i+1}\,x 
\]
hence~$\phi_i\,x$ and $G\,\phi_i\,x$ have $\theta_{i+1}\,x $ as an upper bound and the least upper bound will exist.
\[
\phi_{i+1}\; x \qleq \theta_{i+1}\; x
\]
\lemma{Lemma.}
Define the sequence $(\psi_i)$, where
\[
\psi_0 = \phi_n
\mbox{~~ and ~~}
\psi_{i+1} = G\,\psi_i
\]
The sequence $(\psi_i)$ is well-defined and forms an increasing chain with
$\psi_i\qleq\theta_{i+n}$.

\lemma{Proof.} The proof is by induction as above.
The base case holds trivially and in the induction step we have 
\[
\psi_{i+1}= G\;\psi_i\qleq G\;\theta_{i+n}= \theta_{i+n+1}
\]

\lemma{Lemma.} $\theta_i \qleq \psi_i$.

\lemma{Proof.} The proof is by induction. In the induction step we have 
\[
\theta_{i+1}= G\;\theta_i\qleq G\;\psi_{i}= \psi_{i+1}
\]

\lemma{Lemma.} $\Lub_i \theta_i = \Lub_i \psi_i$.

\lemma{Theorem.} For a sequence $(\phi_i)$ which has stabilised after $n$ iterations we have:
\[  
  \forall x \in S_n. \phi_n\;x = \lfp\; G\; x 
\]
\lemma{Proof.}
By induction we prove that $\forall y\in S_n.\phi_n\,y = \psi_i\, y$
for all $i$. 
By definition $\phi_n = \psi_0$.
Assume that for some $i$ that
$\forall y\in S_n.\phi_n\,y = \psi_i\, y$ and given $x\in S_n$
\[
  \psi_{i+1}\;x = \psi_i\; x\;\lub\; G\;\psi_i\;x = \phi_n\;x
\]
since
\moveup
\[
  \psi_i\;x &=\phi_n\;x
\\
  G\;\psi_i\;x &= G\;\phi_n\;x =\phi_n\;x
\]
In the last step we used the externalised view of neededness. Since $\phi_n$ and $\psi_i$ return the same value for needed arguments then evaluating $G$ with any of these functions will return the same result.

\lemma{Comment.}

The theorem
holds independently of the evaluation strategy. The traditional approach in fixpoint iteration is to use a breadth-first strategy. But it is also possible to use depth-first or a worklist to schedule re-evaluation.

In a fixpoint iteration some values are needed early on but not when the iteration has stabilised. Such values are called spurious arguments. The theorem shows that it is safe to omit those when they are no longer needed.

\lemma{Termination.}
The theorem above states a partial correctness of certain types of fixpoint iteration. There are various ways to guarantee that the iteration terminates. The easiest is to require the domains $D_1$ and $D_2$ to be finite and that the sequence $(S_i)$ is a chain. 

Fixpoint iteration may be used as a general programming paradigm just as recursion or while-loops. When a program uses recursion we may be able to verify that it terminates but as a programming construct it will not in general terminate. We know that there are two sources of non-termination with recursion: circularity and an infinite set of dependent values. This could be exemplified by the functions $f(x)=f(x)$ and $f(x)=f(x+1)$. Similarly there are two sources of non-termination with fixpoint iteration: an infinite set of dependent values and an infinite chain of approximations of the result. This is exemplified by the functions $f(x) = f(x+1)$ over a flat domain of numbers and $f(x)=1+f(x)$ over a domain of numbers with integer ordering and thus of infinite height. The function $f(x)=f(x)$, interpreted over a domain, is easy to implement as it should just return the bottom element of the domain.

\paragraph{Higher-order fixpoint iteration.}
 The higher-order fixpoint operation is based on these principles: In the fixpoint iteration we use the least upper bound of the previous approximation and the re-evaluated value. When we evaluate a function the result only depends on the needed parts of the arguments. This will be developed in the rest of the paper.


\section{An abstract datatype for domains}
\label{sec:imp}

Using 
fixpoint iteration in practice requires efficient representation of values and function graphs. This is even more important in the higher-order case where arguments to functions may be functions or function graphs. In this section we describe a datatype for domains and in the next section how to use it for fixpoint iteration.

We will use Standard ML as implementation language but we only use constructions that are well-known in many programming languages. Values can be represented in a similar way in the Collections Framework in Java. 

Given a domain $D$ we assume that all but limit points may be represented uniquely in a programming language using a type {\ttt D}. 
We further assume that we can provide the following values and operations.
\begin{itemize}
\item A bottom value: {\ttt bot}
\item A less-or-equal operation: {\ttt leq}
\item A least upper bound operation: {\ttt lub}. This operation is not necesarily defined for all arguments since we only require that any two values with an upper bound has a least upper bound. The operation

\item A total order: {\ttt cmp}. This ordering is only used for implementation purposes and does not need to agree with the partial ordering of the domain. Equality in the two orderings must, however, agree since we require values in $D$ to be represented uniquely. The ordering is implemented with an operation that returns a value in the set $\{${\ttt LESS, EQUAL, GREATER}$\}$. Alternatively one could have used values -1, 0, and 1 as often done in the Java/C world.
\end{itemize}

All in all a domain is a set of values plus a tuple of the following operations.
\begin{quote}\begin{verbatim}
datatype 'a domain = 
   Dom of 'a                (* bottom *)
        * ('a*'a -> bool)   (* lessOrEq *)
        * ('a*'a -> bool)   (* equal *)
        * ('a*'a -> order)  (* compareTo *)
        * ('a*'a -> 'a)     (* lub *)
        * ('a -> string);   (* toString *)
\end{verbatim}\end{quote}

From base domains we may construct more complex domains.

\lemma{String domain: \strdom.}
 We can define  \verb'strdom: string domain' as the flat domain of strings. As bottom element we will use the empty string, and the ordering just specifies that the bottom element is less than or equal to all values in the domain.

\lemma{Integer domain: \intdom.}

The set of natural numbers can be made into a domain using the usual integer ordering. We then get the domain $\Int_0^\infty$ of natural numbers with zero as bottom element and infinity as limit point. The infinity value is not represented in the datatype as it is a limit point and will typically be the result of an infinite fixpoint iteration.

\begin{quote}\sss\begin{verbatim}
val bot = 0;
fun leq(a,b:int) = a < b;
fun lub(a,b) = if a < b then b else a;
fun cmp(a,b:int) 
    = if a = b then EQUAL else if a < b then LESS else GREATER;
\end{verbatim}\end{quote}

\lemma{List domain.}
From any domain we can construct the domain of lists of values from that domain. The empty list is the bottom element and shorter lists are less than or equal to longer lists provided that they are related element-wise.
\begin{quote}\begin{verbatim}
listdom: 'a domain -> 'a list domain
\end{verbatim}\end{quote}

\lemma{Tuples}
Two domains can be tupled into a domain of tuples \verb'tupdom'. The bottom element is the tuple of the bottom element from the two domains.
\begin{quote}\begin{verbatim}
tupdom: ('a domain *'b domain) -> ('a *'b) domain
\end{verbatim}\end{quote}

\lemma{Power set domain.}
For a domain we can construct the domain of subsets of values from that domain. The empty set is the bottom element and we use ordinary subset ordering. Sets are represented as sorted lists where elements are sorted according to the total ordering on the domain.
\begin{quote}\begin{verbatim}
setdom: 'a domain -> 'a list domain
\end{verbatim}\end{quote}
Least upper bound is set union and can be implemented as a merge of two sorted lists.  On sets we also define  a \verb'member', intersection and a set difference function.  Notice that the singleton set operation and the set difference function in general are not monotonic. 

We could also construct a set domain using a binary tree representation but we seem to use least upper bounds more frequently than membership tests and in that context the list representation is quite acceptable.

\lemma{Domain of partial function graphs.}
For domains $D_1$ and $D_2$ we can construct the domain of partial function graphs from $D_1$ to $D_2$. A function graph is just a tabulation of a function and it is a way to represent a function as a first-order value. This domain is essentially just the powerset domain of tuples of values from $D_1$ and $D_2$. Since we tend to use lookups more frequently than least upper bounds of function graphs, a representation using binary trees seems preferable. In our implementation language this can be achieved with a special dictionary structure. We could, however, equally well have used a list of tuples.

\begin{quote}\begin{verbatim}
grfdom: ('a domain *'b domain) -> ('a *'b) dict domain
\end{verbatim}\end{quote}

We will also define some extra functions of function graphs:
\begin{quote}\begin{verbatim}
lookupg: ('a *'b) dict -> 'a -> 'b
isdef:   ('a *'b) dict -> 'a -> boolean
updateg: ('a *'b) dict -> ('a,'b) -> ('a *'b) dict
\end{verbatim}\end{quote}
This domain does not use the ordering on $D_1$ and is the domain of all functions from the set $D_1$ to $D_2$. One can also define the domain of all continuous functions from $D_1$ to $D_2$ but that is not what we will use here.

The \verb'lookupg' function will look-up an argument value in a graph and return the result value, if present. Otherwise it returns the bottom element. The function \verb'isdef' checks whether an argument is defined in the graph. The \verb'updateg' updates a graph with a new argument-result pair.


\section{A domain of higher-order function graphs}

The idea in using function graphs to represent functions is to tabulate functions only for needed arguments. In the higher-order case this is a special challenge since we should only tabulate functional arguments to functions for the needed arguments. Consider a function
\[
   g\; x\; k &= k(0) \sqcup k(x)
\\
  m \; x &= 1 \sqcup x
\\
  f\; x &= g\; x\; m
\]
and an initial call $f\; 1$. The function $m$ is not called directly but since a call to $g$ where the first argument is $1$ will call the second argument with $0$ and $1$ then we should tabulate $m$ for those values.

Calls through arguments can also be
described as an analysis of which {\em parts} of an argument
will be used. If an argument is a function we may only need
to call this function with a small set of arguments,
hence only using a small part of this function. The purpose of
this section is to establish a representation of this set.

The central idea is to tabulate arguments for the needed parts when it is used as an argument and not when the argument is used as a function. The problem in this is to specify which parts of arguments will be needed prior to the call. What is needed of the arguments may both depend on the function and the arguments to the function. In this way we have an extra level of circularity in the higher-order case.

Since partially applied functions can appear nested inside
expressions we will use a uniform way to record needs
throughout the type structure.  To record 
which parts of the arguments to a functional value will be
used during evaluation we extend the domain with an extra
part that describes needs.

Let $V_t$ be the usual set we would use to describe values of type $t$. We will now instrument the description with extra need information in a domain $V'_t$
\[
   V_{t1*...*tn \fun tm} = V_{t1}* .. *V_{tn} \fun V_{tm}
\\
   V_{base} = \mbox{ .. numbers and strings}   
\\[\smallskipamount]
   V'_{t1*...*tn \fun tm} = V'_{t1}* .. * V'_{tn} 
    \fun (V'_{tm} *N_{t1} *...* N_{tn} )
\\
   V'_{base} = V_{base} * \Unit
\\  
  N_{t1*...*tn \fun tm} = \power( V'_{t1} * ...* V'_{tn} )
\\
  N_{base} = \Unit
\]
Here $V_{base}$ is a domain of base values (e.g. numbers and strings) and $\Unit$ a one-point domain.


\lemma{The Res domain.}

We will now define a domain of higher-order function graphs. In the higher-order case the argument needs of functions should be an externalised property. For now we only consider how this should be represented; in the next section we will show how they are computed.

We will use a datatype {\ttt Res} to describe these instrumented values. They are either simple values or function graphs from {\ttt Res list} to 
tuples of {\ttt Res} values and needs described as {\ttt Res list list list}.
\begin{quote}\sss\begin{verbatim}
datatype Res = S1 of string | I1 of int | Bot1
             | G1 of ResGraph
withtype ResGraph = (Res list,Res*Res list list list) dict;
type Needs=Res list list list;
\end{verbatim}\end{quote}

A function is represented as a map of arguments (\verb'Res list') to a tuple of a result and a list of sets of argument tuples that arguments (\verb'Res list list list') are called with.

The datatype \verb'Res' can be equipped with both a partial order and a total ordering for implementation purposes, a bottom element \verb'bot_r'
and a least upper bound operation \verb'lub_r' using the domain constructions described in section \ref{sec:imp}. Similarly the types {\ttt Needs}  can be given a domain structure with bottom element \verb'bot_n' and least upper bound \verb'lub_n'.

The type {\ttt ResGraph} will represent function graphs with extra need information and will have the operations defined for graph domains described in section \ref{sec:imp}:
\begin{quote}\sss\begin{verbatim}
bot_g : ResGraph
lub_g : ResGraph * ResGraph
isdef : ResGraph -> (Res list) -> boolean
lookupg : ResGraph -> (Res list) -> (Res * Needs)
updateg : ResGraph -> (Res list,Res * Needs) -> ResGraph
\end{verbatim}\end{quote}

\lemma{Higher-order values.}
The second part of the representation of values is to include the extra graph information into ordinary values in the higher-order functions. We will use a slightly modified definition of the type {\ttt V} compare to the sets $V_t$ shown above.

\begin{quote}\sss\begin{verbatim}
datatype V = S of string | I of int | Bot 
  | C of ResGraph*(V list -> V);
\end{verbatim}\end{quote}

We can establish a close relationship between the \verb'V' set and the \verb'Res' domain with two functions
\verb'v2r : V -> Res' and
\verb'r2v : Res -> V'.

\begin{quote}\sss\begin{verbatim}
fun v2r (S s)    = S1 s
 |  v2r (I i)    = I1 i
 |  v2r Bot      = Bot1
 |  v2r (C(g,_)) = G1 g;

fun r2v (S1 s)   = S s
 |  r2v (I1 i)   = I i
 |  r2v Bot1     = Bot
 |  r2v (G1 g)   = C(g,fn xs => r2v(#1(lookupg g (map v2r xs)));
\end{verbatim}\end{quote}
We may note that for $r$ in \verb'Res' that $r$ = \verb'v2r(r2v('$r$\verb'))' but for 
$v$ in \verb'V' that 
$v \qgeq \,\mbox{\ttt r2v(v2r(}v\mbox{\ttt))}$.

At the outer level we can externalise argument needs with a function 
{\ttt callneed}. It will memoize functions in the argument list, perform the function call and return the result together with the memo information. 
\begin{quote}\sss\begin{verbatim}
callneed : (V list -> V) -> V list -> V * Needs
\end{verbatim}\end{quote}
From a need set we may tabulate functions in a list for the needed parts. Such a function will have type
\begin{quote}\sss\begin{verbatim}
tabulate : V list -> Needs -> Res list
\end{verbatim}\end{quote}
Both these functions are fairly straightforward and will not be described further.
 


\section{Higher-order iteration}

We are now ready to construct the higher-order fixpoint iterator.  We have two graphs {\ttt phi2} and {\ttt phi1} of evaluated values from the current and the previous iteration. The iteration continues until stability has been achieved.

There is an extra level of iteration involved since the set of needed parts of arguments is defined circularly. Initially we assume that no part of functional arguments is needed. We will then attempt to evaluate the function and record which parts of the arguments were needed. The arguments are then tabulated for these parts, and this continues as long as more needs are recorded.

The fixpoint operation has the type:
\begin{quote}\sss\begin{verbatim}
Fix: ((V list-> V)-> V list-> V)-> V list-> V
\end{verbatim}\end{quote}
Internally it uses four functions
\begin{quote}\sss\begin{verbatim}
FF: (V list-> V) -> Res list-> V * Needs
gg: Needs -> V list -> V * Needs
ff: V list -> V 
iterate: V list-> V
\end{verbatim}\end{quote}
It uses the functions from the previous section plus a few extra functions
\begin{quote}\sss\begin{verbatim}
fun fstcall (C(g,f)::rs) = f(rs) | fstcall _ = Bot ;
fun lub_v(x,y) = r2v (lub_r(v2r x,v2r y)); 
fun lub_rn ((v1,n1),(v2,n2)) = (lub_r(v1,v2),lub_n(n1,n2));
fun rs2v rs = map r2v rs;
fun isBot1 (Bot1::xs) = true | isBot1 [] = true | isBot1 _ = false;
fun isC1 ((G1 _)::xs) = true | isC1 _ = false;
\end{verbatim}\end{quote}
The evaluation of functional arguments to functions is done during the tabulation of the arguments.

\begin{quote}\sss\begin{verbatim}
fun Fix F  =
  let val phi1 = ref bot_g; val phi2 = ref bot_g
    fun FF ff rs =
      if isBot1 rs then (Bot,bot_n) else
      if isC1   rs then callneed fstcall (rs2v rs) else 
      if isdef (!phi2) rs then 
        let val (r1,n2)=lookupg (!phi2) rs in (r2v r1,n2) end 
      else
       let val _ = phi2 := update (!phi2) (rs,lookupg(!phi1) rs);
           val (v1,nds) = callneed (F ff) (rs2v rs)
           val r2 = lookupg (!phi2) rs
       in  phi2 := update (!phi2) (rs,lub_rn(r2,(v2r v1,nds)));
          (v1,nds) end;
    fun gg nd vs =
       let val (v1,n1) = FF ff (tabulate vs nd) 
           val n2 = lub_n (n1,nd)
       in if eq_n(nd,n2) then (v1,n2) else gg n2 vs end
    and ff ws = let val (v,n)= gg bot_n ws in v end
    fun iterate xs =
      let val _ = ( phi1:=(!phi2); phi2 := bot_g ); 
          val v1 = ff xs ; 
      in if eq_g(!phi2,!phi1) then v1 else 
         iterate xs 
      end;
  in iterate end ;
\end{verbatim}\end{quote}
The function {\ttt iterate} will re-evaluate a call and recursively needed argument until stability. It uses the the functions {\ttt ff} and {\ttt gg} to tabulate arguments to a function based on which parts of its arguments a function needs. The function {\ttt FF} evaluates other function calls  in a depth-first style.


\lemma{Example: A higher-order fixpoint.}

To use the fixpoint iteration we should express a functional over the domain \verb'V List -> V'. We will here rexamine the function $g'$ from the example with closure-based fixpoint iteration in section \ref{sec:hofi}.
\begin{quote}\sss\begin{verbatim}
fun F call [S "g",n,k] = glb(n,lub(call [k,I 1],call [S "g",n,call [S "m",n,k]]))
  | F call [S "m",n,k,x] = call[k,glb(n,x)]
  | F call [S "top",x] = I 1
  | F call [S "bot",x] = I 0
  | F call [S "fb",x] = call [S "g",x,call [S "bot"]]
  | F call [S "ft",x] = call [S "g",x,call [S "top"]]
  | F call xs = C(bot_g,fn ys => call(xs@ys))
and lub xy = lub_v xy
and glb(I x,I y) = if x > y then I y else I x | glb _ = Bot;
val f =  Fix F;
val rr = f [S "ft", I 1]; 
\end{verbatim}\end{quote}

\[
  \mathit{ft}: & [1] &=> (1,[])
 \\
   g: & [1,[1\mapsto 1]] & =>(1,[[],[[1]]])
 \\
   m: & [1,[1\mapsto 1],1] &=> (1,[[],[[1]],[]])
 \\
 \mathit{top}: & [1] => (1,[])
\]

This should be read as follows. If $g$ is called with $1$ and a function that map $1$ to $1$ then it will return $1$ and call its second argument with $1$. If $m$  is called with $1$, a function that maps $1$ to $1$ and $1$ then it will return $1$ and call its second arguemnt with $1$.

\section{Memoization and second-order fixpoint iteration}

The higher-order fixpoint iteration may be seen as a generalisation of techniques for memoizing functions. Memoization is a well-known technique for avoiding re-evaluation of repeated calls of functions. It requires a certain amount of administrative overhead and is in consequence not normally worth the extra effort. 

Let us start with the Fibonacci function written in Standard ML.
\begin{quote}\begin{verbatim}
fun fib x = if x < 2 then 1 
            else (fib (x-1)) + (fib (x-2));
\end{verbatim}\end{quote}
A call to the function may be written as {\verb'fib 12'} but the function is impractical when used with larger arguments due to re-evaluation of smaller Fibonacci values. 

One approach could be to memoize the Fibonacci function. A simple memoizing transformation of a first order function might look as follows
%
\begin{quote}\begin{verbatim}
fun memo f =
  let val l = ref []
      fun eval [] x 
          = let val r = f x 
            in l := (x,r)::(!l); r end
        | eval ((a,b)::t) x 
          = if x=a then b else eval t x;
    fun ff x = val (!l) x
  in ff end;
val mfib = memo fib;
\end{verbatim}\end{quote}
The function could be called as before and repeated calls will now not result in re-evaluation. It is, however, only the external calls which will be memoized so it is still impractical to use for larger arguments.

If we want to memoize the internal calls the first step is to define the Fibonacci function as a fixpoint of a second-order function.
\begin{quote}\begin{verbatim}
fun Fib f x = if x < 2 then 1 
              else (f (x-1)) + (f (x-2));
fun fix F x = (F (fix F)) x;
val ffib = fix Fib;
\end{verbatim}\end{quote}
The Fibonacci function defined in this way will behave just as the original definition. The fixpoint operation may, however, be changed into a memoizing operation. The fixpoint operation is then
%
\begin{quote}\begin{verbatim}
fun memoFix F xx =
  let val l = ref []
      fun eval [] x = let val r = F ff x 
                      in l := (x,r)::(!l); r end
        | eval ((a,b)::t) x = if x=a then b else eval t x
    and ff x = eval (!l) x
  in ff xx end;
val mmfib = memoFix Fib;
\end{verbatim}\end{quote}
This Fibonacci function will tabulate its internal calls and run in almost linear time if we used an efficient representation of argument-result pairs.

\lemma{Truncated depth-first evaluation.}
The second-order fixpoint operator may be seen as a generalisation of the memoization of second-order functionals and as a special case of the higher-order fixpoint operator. In the second-order case we do not need to iteratively tabulate arguments to functions for needed parts since arguments are of base type.


We call the evaluation strategy for ``truncated depth-first'' since it evaluates in a depth-first style as long as no circularities are detected. When we encounter circularities we truncate the evaluation and just use previously evaluated values or the bottom element instead.

Let $D_1$ and $D_2$ be two domains and let $D_3$ be the graph domain from $D_1$ to $D_2$. Let us further assume that we have defined the functions {\ttt lookupg, isdef,} and {\ttt update} on the graph domain (see section \ref{sec:imp}), let {\ttt lub} be the least upper bound operation on $D_2$,  and {\ttt bot\_g}, {\ttt eq\_g} and {\ttt lub\_g} is the bottom element, equality and least upper bound operations on the graph domain $D_3$. We will here present the fixpoint operation for fixed domains $D_1$ and $D_2$ but in practice it may be parameterised on the domains so that the same function can be used for any domain.
The function should have type
\begin{quote}\begin{verbatim}
Fix: 'a domain -> 'b domain -> (('a -> 'b) -> 'a -> 'b) -> 'a -> 'b
\end{verbatim}\end{quote}
but we will here leave out the domain arguments and assume the various function on domains are defined with the appropriate type as described above:
\begin{quote}\sss\begin{verbatim}
fun Fix F = 
  let val phi1 = ref bot_g (* previous *)
      val phi2 = ref bot_g (* current *)
      fun f x =
        if isdef(!phi2) x then lookupg (!phi2) x else
        let val r0 = lookupg (!phi1) x
            val _ = phi2:= update(!phi2) (x,r0)
            val r = F f x
            val _ = phi2:= update(!phi2) (x,lub(r,r0)) 
        in r end;
      fun iterate x = (
        phi1 := !phi2;  phi2 := bot_g; f x;
        if eq_g(!phi1,!phi2) then lookupg (!phi2) x
        else iterate x);
   in iterate end;
\end{verbatim}\end{quote}

\lemma{Discussion.}

The iteration uses two function graphs: {\ttt phi1} and {\ttt phi2} where {\ttt phi1} contains values found in the last iteration and new values are stored in {\ttt phi2}. The need sets are represented implicitly in the function graphs as the set of defined arguments.


There is no guarantee that users of this function provide a correct pseudo-monotonic function. If the {\ttt lub} does not exist then it must be because of errors in the function {\ttt F}. Similarly if the iteration does not stop then it is because the iterated needs are infinite.

The truncated depth-first evaluation strategy has been used in a number of different versions in program analysis, circular attribute grammars etc. over the years \cite{mr:phd,f:agfix}. It is both simple and more efficient than the breadth-first evaluation strategy often referred to as minimal function graphs. 

The approach here will automatically remove spurious calls \cite{mr:hcfi,B335}. In the stable situation the graph {\ttt phi2} will be tabulated for all arguments which directly or indirectly are needed from the initial call - except for those which are already in the graph of found fixpoints.

\lemma{Variations.}

An obvious extension is to memoize the fixpoint function so that repeated calls will not require new iteration. The memoization can easily be built into the fixpoint function since argument-result pairs already are stored in a graph.

The iteration stops when re-evaluation of the argument and all dependent arguments do not result in changed values. This also means that the iteration is at least done twice. If the dependency is not circular we will do an unnecessary re-evaluation. An alternative approach could be to store used values during evaluation in a separate graph and to re-evaluate until the used-value graph is a subgraph of the computed graph in the iteration. This method will be optimal for programs without circular dependencies but at the cost of some extra book-keeping. Whether it is worthwhile depends on the function used as argument to the fixpoint.

When the iteration is truncated because of circularities we will use the value from the previous iteration or the bottom element. In some domains it could be relevant instead to use the least upper bound of all values in the graph where the arguments are less than or equal to the one we are searching for. Such an operation is quite costly but may be worth it if the argument domain has a complex structure.

\lemma{Dependency-based methods.}
 
Over the last two decades a number of authors have proposed techniques which store a dependency graph during the iteration. The idea is to restrict re-evaluations so that they only occur when the graph has been changed for some of the arguments it depends on. 
The top-down evaluator \cite{lch:genfix} evaluates in a depth-first order but with local computation of fixpoints. The neededness analysis method \cite{nj:need} uses breadth-first strategy, but limits the re-evaluation to arguments that depend on arguments that have been changed. The time stamp solver WRT \cite{fs:evenfast} uses extra time stamps in the work list to schedule re-evaluation in a fair manner.

There is a significant cost in maintaining dependency information and any efficiency gain requires that each evaluation of the function {\tt F} is costly.

\section{Examples}

In this example we will need a second-order fixpoint operator which has the type

\begin{quote}\begin{verbatim}
Fix: 'a domain -> 'b domain -> (('a -> 'b) -> 'a -> 'b) -> 'a -> 'b
\end{verbatim}\end{quote}
We will now give two examples of the programming style one may use with this fixpoint operator. 

\lemma{First-order strictness analysis.}

%

This example is a strictness analyser for a small first-order functional language. We assume programs are lists of function definitions. A function definition consists of a function name, a parameter list, and an expression.

\begin{quote}\begin{verbatim}
datatype exp = 
  par of string | cst of int | 
  add of exp * exp | 
  cond of exp * exp * exp | 
  call of string * exp list;
type progs = (string*(string list * exp)) list;
\end{verbatim}\end{quote}

The strictness analyser is defined as the fixpoint of a second-order function.
\begin{quote}\begin{verbatim}
fun F prg seval (f,args) =
   let fun seval1 env (cst i) = 1
        |  seval1 env (par s) = lookup s env
        |  seval1 env (add (e1,e2)) = 
             glb(seval1 env e1,seval1 env e2)
        |  seval1 env (cond (e0,e1,e2)) =
             glb(seval1 env e0,
              lub(seval1 env e1,seval1 env e2))
        |  seval1 env (call (f,exps)) =
             seval (f,map (seval1 env) exps);
       val (pars,exp) = lookup f prg 
   in seval1 (zip pars args) exp end;
fun Strict p 
  = Fix (tupdom (strdom,listdom intdom)) 
        intdom (F p);
\end{verbatim}\end{quote}
We have here used a \verb'zip' function to collapse two lists into an association list, and \verb'lookup' to lookup values in an environment.


For a program \verb'p' which contains a function \verb'f' with two parameters
the strictness analysis may be performed as follows.
\begin{quote}\begin{verbatim}
val strict = Strict p;
strict ("f",[0,1]);  strict ("f",[1,0]);
\end{verbatim}\end{quote}
The function \verb'strict' is now a function of type \verb'string*int' \verb'list -> int'. The domain structure is hidden as soon as we have defined how the fixpoint should be computed and the function may be used as an ordinary function in the language. Notice that we do not define the function as
\begin{quote}\begin{verbatim}
fun strict args = Strict p args;
\end{verbatim}\end{quote}
If we used the latter definition every new call to \verb'strict' would re-evaluate the fixpoint rather than using the built-in memoization in the fixpoint operator.

\paragraph*{First-set for grammars.}

For a non-terminal in a context-free grammar the First-set is the set of terminals it may generate. The First and Follow sets for context-free grammars play an important role in constructing parsers and they are normally defined as least fixpoints.

In this example a grammar is a list of productions with a string as left-hand-side and a list of terminals or non-terminals as right-hand side:
\begin{quote}\begin{verbatim}
datatype elem = nt of string | tm of string ;
type gram = (string * elem list) list;
\end{verbatim}\end{quote}
The first-function can now be defined as the least fixpoint of the following function.

\begin{quote}\begin{verbatim}
fun F g first s = firstg first s g
and firstg first s [] = []
 |  firstg first s ((nt1,rh)::r) = 
      union(if s=nt1 then firstrh first rh 
            else [], firstg first s r)
and firstrh first [] = [""] 
 |  firstrh first ((nt s)::rr) = 
       let val ff = first s 
       in union(ff, if member "" ff 
             then firstrh first rr else [])
       end
 |  firstrh first ((tm s)::_) = [s];
fun First g  = Fix strdom (setdom strdom) (F g);
\end{verbatim}\end{quote}
Let us now consider a small grammar for simple expressions.

\begin{quote}\begin{verbatim}
val grm = 
  [("exp",   [nt("term")]),
   ("exp",   [nt("exp"),tm("+"),nt("term")]),
   ("term",  [nt("factor")]),
   ("term",  [nt("term"),tm("+"),nt("factor")]),
   ("factor",[tm("name")]),
   ("factor",[tm("number")]),
   ("factor",[tm("("),nt("exp"),tm(")")]) ];
val first = First grm;
val r = first "exp";
\end{verbatim}\end{quote}
The whole fixpoint will be tabulated during this call so subsequent calls of \verb'first' for other non-terminals will not result in any re-evaluation.




\section{Discussion and related work}

This paper is concerned with demand-driven solution of
fixpoint iteration. The immediate application is in a
solution of demand-driven versions of program analysis
problems. The fixpoint iteration makes it possible to implement more precise and complex higher-order analyses.
The fixpoint operation is implemented in Standard ML (Moscow ML \cite{mosml}). 


\lemma{Representation techniques.}

Several methods are based on ways of representing the full tabulation
with only a small set of points. 
Amongst these one should mention the 
frontiers algorithm
\cite{hh:front,mh:finlat}
and binary decision diagrams. Such techniques may drastically reduce 
the need for tabulation, though they still seem to require too much 
storage for certain examples. If it is not vital to find the least 
solution but only important to find a safe solution, such methods may 
be connected with 
various approximation techniques in order to improve efficiency. 

\paragraph{Lazy types.}

Our approach seems comparable in efficiency to the work by 
Le Metayer \&{} Hankin on lazy types \cite{lazytype} and the work by Hughes on 
sequential algorithms. 
The precise relationship between such techniques is still not fully established.

\paragraph{Minimal function graphs} 
This paper is concerned with demand-driven solution of
fixpoint iteration. The immediate application is in the
solution of demand-driven versions of (bottom-up) program analysis
problems. If we want to analyse the strictness of a
parameter to a given function then these techniques may
restrict the fixpoint iteration to those parts of the
program that may influence the result.  Chaotic fixpont
iteration \cite{cc:77c} in abstract interpretation is a
method to construct demand-driven versions for analysis problems.
In fact whether an abstract
interpretation should solve a demand or a global dataflow
analysis problem is independent of its specification but
merely a question of how the fixpoint operator is
implemented \cite{cc:77,cc:77c}.

A different problem is the construction of top-down analyses
which, from a description of the possible initial calls, 
examines which values may reach a given part of the program.
For functional programming languages such analyses are often
based on a minimal function graph semantics
\cite{jm:mfg,mr:mfg,mr:hcfi}.

\paragraph*{Third-order fixpoints.} 

Since function graphs can be made into a domain we can tabulate a first order function and use it as a simple finite value. In this way a third order function may be treated as if it were a second-order function and used in fixpoint evaluation as described above. The classical example of strictness analysis of \verb'foldr' is quite easy to analyse in this way. In principle this may be used for even higher-order functions, but it quickly becomes impractical to tabulate functions for all possible arguments.

\paragraph*{Comparison.}

We have implemented a number of different fixpoint operations and compared them for a number of different applications. Below we show some results from calling the \verb'first' function with a grammar for Java and the nonterminal for expressions as arguments. We compare the number of times the functional is called during the fixpoint iteration (\#rhs) and the number of comparisons on values in the string domain (\#cmp)

The fixpoint operations are:
\begin{itemize}
\item Kleene: a simple breadth-first evaluation strategy
\item Dep: a neededness-analysis based evaluator. It is a fairly direct implementation of the evaluator from \cite{nj:need}, though without the monotonic completion which is quite costly.
\item TD: a top-down solver based on \cite{lch:genfix}, though here taken from \cite{fs:evenfast}.
\item W: a worklist solver from \cite{fs:evenfast}.
\item TDF: the truncated-depth-first solver described earlier
\item TDF-sub: the truncated-depth-first solver but with an extra graph of used values. It iterates until the graph has the graph of used values as a subgraph.
\end{itemize}

\begin{tabular}{|l|r|r|}
\hline
Method & \#cmp & \#rhs\\\hline\hline
Kleene& 31352& 572\\\hline
Dep   & 15353& 190\\\hline
TD    & 11377& 66\\\hline
W     & 10413& 147\\\hline
TDF   &  4873& 148\\\hline
TDF-sub& 4331& 111\\\hline
\end{tabular}
\medskip

The strategies fall into three groups
\begin{itemize}
\item The Kleene strategy is not optimized in any way and just iterates until stability. It is naturally more costly than the other techniques. 
\item Dependency-based strategies may be able to use dependency information to determine whether a re-evauation can change the result and thereby avoid some evaluations. The savings in fewer evaluations are at the expense of maintaining dependency information and testing whether dependent values have changed.
\item The truncated depth-first strategy does not maintain dependency information. 
The depth-first approach will normally give fewer iterations but it may re-evaluate some expressions that does not depend on parts that have changed value.
\end{itemize}

In our experience it is quite costly to maintain a dependency graph during fixpoint evaluation. It may avoid some evaluations of the function but at the cost of a certain amount of extra bookkeeping. This may be attractive if each call to the function is very expensive but not otherwise. We have made the same comparison of the fixpoint operation in different program analyses, and in our experience the TDF is both simple and efficient.

\lemma{Logical relations.}
We have not examined correctness proofs of abstract interpretations in this paper. One way to integrate correctness proofs and a standard semantics into the same framework is to base proofs on logical relations and let standard semantics and abstract interpretations follow a similar programming style. This is explored in \cite{mr:phd}.


\section{Conclusion}

We have presented a technique for solving fixpoint equations 
involving higher-order functions. The method may be used for 
higher-order program analysis and more generally as a tool in programming.
For second-order functionals 
{\em needs} are sets of argument tuples. Our fixpoint iteration approach
extends this notion of {\em needs} to the higher-order case 
where {\em needs} are higher-order functionals in arguments and needs.

The fixpoint operator on domains makes it possible to view a class of abstract interpretations as programs and examine the style and properties of the programming paradigm. The work by David Schmidt on denotational semantics led to further analysis of state in functional programs and techniques for identifying global variables in programs. Similarly it would be interesting to see whether more algorithms could be written in a higher-order fixpoint iteration style.


\nocite{ds:dspl}
\nocite{D140}
\nocite{ds:ds}
\nocite{B161}
\nocite{conf/sas/AltM95}
\nocite{conf/esop/CousotCFMMMR05}
\nocite{mh:finlat}

\nocite{*}
\bibliographystyle{eptcs}
\bibliography{dsfest-ref}
\end{document}